\documentclass[twoside,slac_one]{revtex4}
\usepackage{graphicx}
\usepackage{fancyhdr}
\usepackage{amsmath} 
\usepackage{bm}
\usepackage{amsxtra}
\usepackage{amssymb}
\usepackage{amsthm}
\usepackage{latexsym}
\usepackage{lscape}

\begin{document}

\title{\LARGE Geodesically Complete Universe}

%

\author{\Large Itzhak Bars}
\affiliation{Dept of Physics and Astronomy, University of Southern
California, Los Angeles, CA 90089, USA}

\begin{abstract}
This talk is about solving cosmological equations analytically
without approximations, and discovering new phenomena that could not
be noticed with approximate solutions. We found all the solutions of
the Friedmann equations for a specific model, including all the
zero-size-bounce solutions that do not violate the null energy
condition, as well as all the finite-size-bounce solutions, and then
discovered model independent phenomena. Among them is the notion of
geodesic completeness for the geometry of the universe. From this we
learned a few new general lessons for cosmology. Among them is that
anisotropy provides a model independent attractor mechanism to some
specific initial values for cosmological fields, and that there is a
period of antigravity in the history of the universe. The results
are obtained only at the classical gravity level. Effects of quantum
gravity or string theory are unknown, they are not even formulated,
so there are new theoretical challenges.
\end{abstract}

\maketitle

\thispagestyle{fancy}



\section{Introduction}

This work started with the discovery of new techniques for solving
the Friedmann equations analytically \cite{inflationBC}, and finding
the complete set of solutions for the gravitational system that
includes a scalar field $\sigma\left(  x^{\mu}\right)  $ with a
special, but still typical, potential
$V\left(  \sigma\right)  $ in a cosmological model \cite{inflationBC}%
\cite{perturbationCD}. The special techniques were direct outcomes
from
2T-physics \cite{2TPhaseSpace} and 2T-gravity \cite{2Tgravity}%
\cite{2TgravityGeometry}. This led to an emergent formulation of
gravity in 1T-physics with an additional local conformal (Weyl)
symmetry. The cosmological model that emerged in this approach had a
strong overlap with the cosmological colliding branes scenario
\cite{KOSST}\cite{branesMT} developed from worldbrane notions
\cite{RandallSundrum} in M-theory \cite{HoravaWitten}. This
connection led to a fruitful collaboration between our group, that
includes C.H. Chen, and the group that includes Paul Steinhardt and
Neil Turok. Our collaborative work has partially appeared in
\cite{BCT1}. Here I will summarize results that include some that
are about to be published in \cite{BCST2}.

Our approach begins with the standard action typically used in
cosmological models that describe a scalar field $\sigma\left(
x^{\mu}\right)  $ minimally
coupled to gravity%
\begin{equation}
S=\int d^{4}x\sqrt{-g}\left\{
\begin{array}
[c]{c}%
\frac{1}{2\kappa^{2}}R\left(  g\right)
-\frac{1}{2}g^{\mu\nu}\partial_{\mu
}\sigma\partial_{\nu}\sigma-V\left(  \sigma\right) \\
+radiation+matter
\end{array}
\right\}  .\label{action}%
\end{equation}
Although not necessary, the scalar field $\sigma$ could be
identified with the dilaton in string theory, written in the
Einstein frame. For the purpose of studying cosmology, the relevant
degrees of freedom are homogeneous fields that are only time
dependent. The space dependence is ignored in the leading
approximation. It can be argued \cite{Damour} that the space
gradients are responsible for non-leading effects as compared to the
dominant effects parameterized by only time dependent fields that we
study here (including anisotropy).

Our cosmological analysis of this theory involves geometries with or
without spacial curvature, as well as anisotropy, and includes
radiation and other relativistic matter approximated as relativistic
dust described by an energy-momentum tensor consistent with
conformal symmetry. The geometries we
study have the form%
\begin{equation}
ds^{2}=-dt^{2}+a^{2}\left(  t\right)  ds_{3}^{2}=a^{2}\left(
\tau\right)
\left(  -d\tau^{2}+ds_{3}^{2}\right)  ,\label{metric}%
\end{equation}
where $a^{2}\left(  \tau\right)  $ is the scale factor that tracks
the size of the universe. We will use conformal time $\tau$ which is
related to cosmic time $t,$ by $dt=a\left(  \tau\right)  d\tau,$ as
in Eq.(\ref{metric}). The 3-dimensional part $ds_{3}^{2}$ of the
geometries we study include

\begin{itemize}
\item the isotropic FRW universe with or without curvature $K$
\begin{equation}
\left(  ds_{3}^{2}\right)
_{FRW}=\frac{dr^{2}}{1-Kr^{2}}+r^{2}\left(
d\theta^{2}+\sin^{2}\theta d\phi^{2}\right)  ;\;K=\frac{k}{r_{0}^{2}%
},\label{frw}%
\end{equation}

\item the anisotropic Kasner metric with zero curvature
\begin{equation}
\left(  ds_{3}^{2}\right)
_{Kasner}=e^{-2\sqrt{2/3}\kappa\alpha_{1}}\left( dz\right)
^{2}+e^{\sqrt{2/3}\kappa\alpha_{1}}\left(  e^{\sqrt{2}\kappa
\alpha_{2}}\left(  dx\right)
^{2}+e^{-\sqrt{2}\kappa\alpha_{2}}\left(
dy\right)  ^{2}\right)  ,\label{kasner}%
\end{equation}
where $\alpha_{1}\left(  \tau\right)  $ and $\alpha_{2}\left(
\tau\right)  $ are two dynamical degrees of freedom in the metric
along with $a\left( \tau\right)  ,$ and
\end{itemize}

\begin{itemize}
\item the anisotropic Bianchi IX metric with curvature $K$%
\begin{equation}
\left(  ds_{3}^{2}\right)
_{IX}=e^{-2\sqrt{2/3}\kappa\alpha_{1}}\left( d\sigma_{z}\right)
^{2}+e^{\sqrt{2/3}\kappa\alpha_{1}}\left(  e^{\sqrt
{2}\kappa\alpha_{2}}\left(  d\sigma_{x}\right)
^{2}+e^{-\sqrt{2}\kappa
\alpha_{2}}\left(  d\sigma_{y}\right)  ^{2}\right)  ,\label{bianchi9}%
\end{equation}
where $d\sigma_{i},$ which are given in \cite{Misner}, depend on
space coordinates (and $K$) but not on $\tau.$ The Bianchi IX metric
reduces to the FRW metric with curvature $K$ in the isotropic limit
$\alpha_{i}\rightarrow0,$ and it reduces to the anisotropic Kasner
metric when the curvature vanishes $K\rightarrow0.$
\end{itemize}

The relevant degrees of freedom are $\sigma\left(  \tau\right)
,a_{E}\left( \tau\right)  ,\alpha_{1}\left(  \tau\right)
,\alpha_{2}\left(  \tau\right) .$ Here we have attached the label
$E$ on the scale factor $a_{E}\left( \tau\right)  $ to emphasize
that it is defined in the Einstein frame. We do this because we will
also define other frames to solve the equations. The equations of
motion (namely the Friedmann-type equations) are obtained from the
action above. However, it is convenient to provide an effective
worldline mechanics-type action in the Einstein frame
$S_{\text{eff}}^{E}$ that
reproduces the same equations%
\begin{equation}
S_{\text{eff}}^{E}=\int d\tau\left\{
\begin{array}
[c]{c}%
\frac{1}{e}\left[  -\frac{6}{2\kappa^{2}}\dot{a}_{E}^{2}+\frac{1}{2}a_{E}%
^{2}\dot{\sigma}^{2}+\frac{1}{2}a_{E}^{2}\dot{\alpha}_{1}^{2}+\frac{1}{2}%
a_{E}^{2}\dot{\alpha}_{2}^{2}\right] \\
-e\left[  a_{E}^{4}V\left(  \sigma\right)  +\rho_{0}-\frac{6K}{2\kappa^{2}%
}a_{E}^{2}v\left(  \alpha_{1},\alpha_{2}\right)  \right]
\end{array}
\right\}  .\label{SeffE}%
\end{equation}
Here the constant $\rho_{0}$ represents the energy density of
radiation and relativistic matter. It comes from approximating these
degrees of freedom as ``dust'' modeled as a perfect fluid, with an
energy-momentum tensor of the form $T_{\mu\nu}=\left(  \rho+p\right)
u_{\mu}u_{\nu}+pg_{\mu\nu},$ that is taken to be traceless
($p=3\rho$) and behave as $\rho\left(  \tau\right)
=\rho_{0}/a_{E}^{4}\left(  \tau\right)  $ to be consistent with the
conformal symmetry for massless particles (including radiation and
matter). The
potential energy term for the anisotropy involving $v\left(  \alpha_{1}%
,\alpha_{2}\right)  $
\begin{equation}
v\left(  \alpha_{1},\alpha_{2}\right)  =\frac{1}{3}\left(
-e^{-4\sqrt
{2/3}\kappa\alpha_{1}}-4e^{2\sqrt{2/3}\kappa\alpha_{1}}\sinh^{2}\left(
\sqrt{2}\kappa\alpha_{2}\right)
+4e^{-\sqrt{2/3}\kappa\alpha_{1}}\cosh\left(
\sqrt{2}\kappa\alpha_{2}\right)  \right)
\end{equation}
is obtained by computing $R\left(  g\right)  $ for the Bianchi IX
metric \cite{Misner}. The additional degree of freedom $e\left(
\tau\right)  $ plays the role of the metric on the worldline and
ensures $\tau$-reparameterization invariance. This effective action
can be derived directly from the 4-dimensional action by inserting
the metric including the lapse function $ds^{2}=a_{E}^{2}\left(
\tau\right)  \left(  -d\tau^{2}e^{2}\left( \tau\right)
+ds_{3}^{2}\right)  ,$ where the lapse function $e\left( \tau\right)
$ is the einbein on the worldline. The equation of motion for
$e\left(  \tau\right)  $ imposes the zero energy constraint which is
equivalent to the $G_{00}=T_{00}$ Einstein equation. Eventually we
work in the gauge $e\left(  \tau\right)  =1$ which corresponds to
the conformal time in Eq.(\ref{metric}). Similarly, it can be
checked that this effective action reproduces all Friedmann-type
equations that are derived from the original action in
Eq.(\ref{action}).

For our model we chose the following potential energy
\begin{equation}
V\left(  \sigma\right)  =\left(  \frac{6}{\kappa^{2}}\right)
^{2}\left( c\sinh^{4}\left(
\sqrt{\frac{\kappa^{2}}{6}}\sigma\right)  +b\cosh^{4}\left(
\sqrt{\frac{\kappa^{2}}{6}}\sigma\right)  \right)  ,\label{V}%
\end{equation}
After insuring stability by requiring $\left(  b+c\right)  >0,$ the
plot of this potential, which looks like a well or double well for
various values and signs of $b,c,$ shows that it has features that
are similar to other potentials used in cosmological applications.
The reason for choosing this specific potential is that we can use
some special tricks to solve the Friedmann equations exactly in the
FRW universe and find all of the solutions analytically
\cite{inflationBC}\cite{BCT1}\cite{BCST2}. We can similarly solve
the equations analytically for a few other potentials, as presented
in a future publication.

\section{BCST Transformation and Weyl Symmetry}

The tricks for solving the equations are actually deeply related to
the gauge
symmetries and shadow phenomena in 2T-physics \cite{2TgravityGeometry}%
\cite{inflationBC}. But, in the current setting with the potential
$V\left( \sigma\right)  $ in (\ref{V}), it is possible to introduce
them directly as the Bars-Chen-Steinhardt-Turok (BCST)
transformation of the fields $\left( a_{E}^{2},\sigma\right)  $ to a
new basis $\left(  \phi_{\gamma},s_{\gamma }\right)  $ as follows
\begin{equation}
\text{ }a_{E}^{2}=\left\vert z\right\vert ,\;z\equiv\frac{\kappa^{2}}%
{6}\left(  \phi_{\gamma}^{2}-s_{\gamma}^{2}\right)
,\;\;\sigma=\frac{\sqrt
{6}}{4\kappa}\ln\left(  \frac{\left(  \phi_{\gamma}+s_{\gamma}\right)  ^{2}%
}{\left(  \phi_{\gamma}-s_{\gamma}\right)  ^{2}}\right)  .\label{bcst1}%
\end{equation}
Then the action $S_{\text{eff}}^{E}$ in Eq.(\ref{SeffE}) is
rewritten in the new basis as follows
\begin{equation}
S_{\text{eff}}^{\gamma}=\int d\tau\left\{
\begin{array}
[c]{c}%
\frac{1}{e}\left[
-\frac{1}{2}\dot{\phi}_{\gamma}^{2}+\frac{1}{2}\dot
{s}_{\gamma}^{2}+\frac{\kappa^{2}}{12}\left(  \phi_{\gamma}^{2}-s_{\gamma}%
^{2}\right)  \left(
\dot{\alpha}_{1}^{2}+\dot{\alpha}_{2}^{2}\right)  \right]
\\
-e\left[  \phi_{\gamma}^{4}f\left(  s_{\gamma}/\phi_{\gamma}\right)
+\rho _{0}-\frac{K}{2}\left(
\phi_{\gamma}^{2}-s_{\gamma}^{2}\right)  v\left(
\alpha_{1},\alpha_{2}\right)  \right]
\end{array}
\right\}  ,\label{SeffG}%
\end{equation}
where $\phi_{\gamma}^{4}f\left(  s_{\gamma}/\phi_{\gamma}\right)
=b\phi_{\gamma}^{4}+cs_{\gamma}^{4}$ when $V\left(  \sigma\right)  $
is given as in Eq.(\ref{V}).

In the absence of anisotropy, $\alpha_{i}\rightarrow0,$ $v\left(
\alpha _{1},\alpha_{2}\right)  \rightarrow1,$ this form of the
action, with the special potential, shows that the fields $\left(
\phi_{\gamma},s_{\gamma }\right)  $ are decoupled from each other,
except for the zero energy condition that results from the
$e$-equation of motion. The decoupled second order field equations
for $\phi_{\gamma}\left(  \tau\right)  $ and $s_{\gamma }\left(
\tau\right)  $ are easily solved because they are analogous to a
particle in a potential. The solutions are given in terms of Jacobi
elliptic functions. The zero energy condition merely relates the
integration variables $E_{\phi},E_{s}$ (energy levels for
$\phi_{\gamma},s_{\gamma}$) in the form $E_{\phi}=E+\rho_{0}$ and
$E_{s}=E,$ so that there is only one integration variable, $E,$ and
two initial values $\phi_{\gamma}\left(  \tau_{0}\right)
,s_{\gamma}\left(  \tau_{0}\right)  .$ Due to time translation
symmetry one of the initial values has no physical significance;
therefore $E,\phi\left( \tau_{0}\right)  $ are the only two
integration parameters. Together with the 4 parameters of the model
$\left(  b,c,K,\rho_{0}\right)  ,$ these 6 parameters determine the
properties of all the solutions in the absence of anisotropy.

This shows that the BCST transformation makes the Friedmann
equations solvable analytically and leads to the complete set of
solutions given in
\cite{inflationBC}\cite{BCT1}\cite{BCST2} as outlined below.%

\begin{center}
\includegraphics[
height=2.3782in, width=3.5699in
]%
{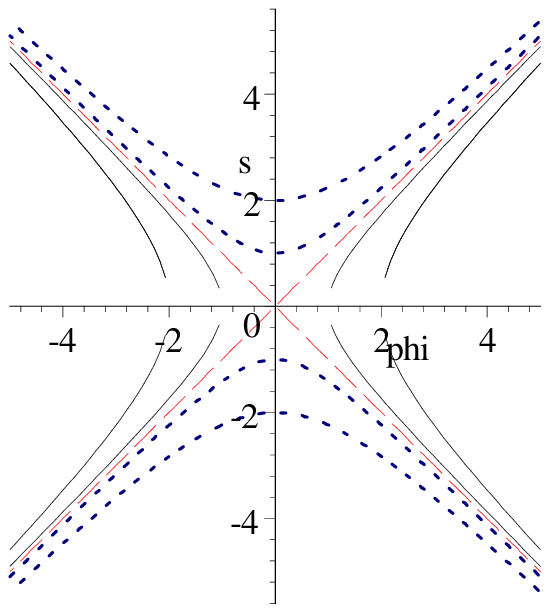}%
\\
Fig.1- Gravity $\phi^{2}>s^{2}$ in left/right quadrants. Antigravity
$\phi ^{2}<s^{2}$ in top/bottom quadrants.
\label{CCrFig1}%
\end{center}

Eq.(\ref{bcst1}) is only half of the BCST transformation between the
$\left( a_{E}^{2},\sigma\right)  $ and $\left(
\phi_{\gamma},s_{\gamma}\right)  $ bases. The form in
Eq.(\ref{bcst1}) is initially defined only for the region of the
$\left(  \phi_{\gamma},s_{\gamma}\right)  $ plane that satisfies
$\left(  \phi_{\gamma}^{2}-s_{\gamma}^{2}\right)  >0,$ which
corresponds to the left and right quadrants as depicted in Fig.1.
The cosmological singularity at $a_{E}^{2}=0$ corresponds to the
45$^{o}$ lines, which may be considered a \textquotedblleft
lightcone\textquotedblright\ in the $\left(
\phi_{\gamma},s_{\gamma}\right)  $ space. The study of geodesic
completeness at the $a_{E}^{2}=0$ singularity amounts to the
question of how geodesics cross the lightcone in $\left(
\phi_{\gamma},s_{\gamma}\right)  $ field
space. Through our explicit analytic solutions \cite{inflationBC}%
\cite{BCT1}\cite{BCST2} we learned that \textit{generic trajectories
in the absence of anisotropy} cross this lightcone and continue to
develop in the entire $\left(  \phi_{\gamma},s_{\gamma}\right)  $
plane. Therefore, geodesic completeness leads us to expand the
domain of the Einstein frame beyond the $\left(
\phi_{\gamma}^{2}-s_{\gamma}^{2}\right)  >0$ region by including the
full $\left(  \phi_{\gamma},s_{\gamma}\right)  $ plane. This
extension of the field domain, including the top and bottom
quadrants, reminds us of a similar extension of spacetime (rather
than field space) via the Kruskal-Szekeres coordinates of the
Schwarzschild blackhole.

To complete the BCST transformation, we also need the inverse
transformation which involves the four quadrants in the $\left(
\phi_{\gamma},s_{\gamma }\right)  $ space. This is given by
\begin{equation}
\phi_{\gamma}=\pm\left\{
\begin{array}
[c]{c}%
\frac{\sqrt{6}}{\kappa}\sqrt{\left\vert z\right\vert }\cosh\left(
\frac{\kappa\sigma}{\sqrt{6}}\right)  ,\text{ if }z>0\\
\frac{\sqrt{6}}{\kappa}\sqrt{\left\vert z\right\vert }\sinh\left(
\frac{\kappa\sigma}{\sqrt{6}}\right)  ,\text{ if }z<0
\end{array}
\right.  ,\;s_{\gamma}=\pm\left\{
\begin{array}
[c]{c}%
\frac{\sqrt{6}}{\kappa}\sqrt{\left\vert z\right\vert }\sinh\left(
\frac{\kappa\sigma}{\sqrt{6}}\right)  ,\text{ if }z>0\\
\frac{\sqrt{6}}{\kappa}\sqrt{\left\vert z\right\vert }\cosh\left(
\frac{\kappa\sigma}{\sqrt{6}}\right)  ,\text{ if }z<0
\end{array}
\right. \label{bcst2}%
\end{equation}
Actually the $\phi_{\gamma}^{2}>s_{\gamma}^{2}$ and $\phi_{\gamma}%
^{2}<s_{\gamma}^{2}$ field domains are separated from each other by
the spacetime singularity at $a_{E}^{2}=0;$ so
$\phi_{\gamma}^{2}>s_{\gamma}^{2}$ is satisfied in a patch of
spacetime, while $\phi_{\gamma}^{2}<s_{\gamma}^{2}$ is satisfied in
a different patch of spacetime. These patches communicate with each
other through the cosmological singularity, or equivalently through
the whole lightcone in the $\left(  \phi_{\gamma},s_{\gamma}\right)
$ plane.

Allowing $\left(  \phi_{\gamma},s_{\gamma}\right)  $ to be extended
to the full $\left(  \phi_{\gamma},s_{\gamma}\right)  $ plane,
including the top and bottom quadrants (as guided by the solutions),
defines a continuation of the Einstein frame to a larger domain of
field space. We found that the full $\left(
\phi_{\gamma},s_{\gamma}\right)  $ provides a geodesically complete
geometry, while the domain $\phi_{\gamma}^{2}>s_{\gamma}^{2}$ by
itself is geodesically incomplete. We find that in the extended
domain the physics described by $S_{\text{eff}}^{\gamma}$ includes
patches of spacetime in which there is gravity when
$\phi_{\gamma}^{2}>s_{\gamma}^{2},$ or antigravity when
$\phi_{\gamma}^{2}<s_{\gamma}^{2}.$ The actions
$S_{\text{eff}}^{\gamma },S_{\text{eff}}^{E}$ are equivalent to each
other via the BCST transformation only in the sector in which
$\left(  \phi_{\gamma}^{2}-s_{\gamma}^{2}\right) $ is positive$.$ In
the antigravity patches the inverse BCST transformation
produces again an Einstein frame, and an action similar to $S_{\text{eff}}%
^{E}$, but with a negative Newton constant.

To better understand the meaning of the different patches we need
the complete theory, not only the effective theory
$S_{\text{eff}}^{\gamma}$ for the cosmological degrees of freedom.
This is given from first principles by a slight extension of the
Einstein action by including only gauge degrees of freedom
associated with a local scaling symmetry, called Weyl symmetry. The
local scaling transformation, given by
\begin{equation}
\left(  \phi\left(  x\right)  ,s\left(  x\right)  \right)
\rightarrow\left( \phi\left(  x\right)  ,s\left(  x\right)  \right)
e^{\lambda\left(  x\right) }\;\text{and\ \ }g_{\mu\nu}\left(
x\right)  \rightarrow g_{\mu\nu}\left( x\right)  e^{-2\lambda\left(
x\right)  },
\end{equation}
does not allow the usual Einstein-Hilbert term, but allows
conformally coupled scalars. The following action contains two
conformally coupled scalars $\phi,s,$ interacting with the curvature
term with the coefficient $\frac {1}{12}$ dictated by the gauge
symmetry
\begin{equation}
S=\int d^{4}x\sqrt{-g}\left(
\frac{1}{2}g^{\mu\nu}\partial_{\mu}\phi
\partial_{\nu}\phi-\frac{1}{2}g^{\mu\nu}\partial_{\mu}s\partial_{\nu}%
s+\frac{1}{12}\left(  \phi^{2}-s^{2}\right)  R\left(  g\right)
-\phi
^{4}f\left(  \frac{s}{\phi}\right)  \right)  .\label{C-action}%
\end{equation}
Here $f(z)$ is an arbitrary function of the scale invariant ratio
$z\equiv\frac{s}{\phi}.$ Fermionic and gauge fields, as well as more
conformally coupled scalars, can be added, as part of a complete
model, without breaking the conformal gauge symmetry
\cite{2TgravityGeometry}. Here $\left(  \phi^{2}-s^{2}\right) /12$
appears as a gravitational parameter that replaces the Newton
constant $1/2\kappa^{2}$. To have the possibility for it to be
positive, we must have one scalar, namely $\phi,$ to have the wrong
sign kinetic term. If $\phi$, like $s,$ has the correct sign, then
we would end up with a purely negative gravitational parameter.
Hence the relative minus sign in $\left(  \phi^{2}-s^{2}\right)  $
is required. The wrong sign kinetic term makes $\phi$ potentially a
ghost. However, the local gauge symmetry compensates for the ghost,
thus insuring unitarity of the theory\footnote{There are other ways
of coupling a second scalar $s,$ beyond $\phi,$ consistently with
the Weyl symmetry \cite{2tSugra}. Only the form given here parallels
the structure $(\phi, s)$ of the complete set of solutions. Other
schemes of introducing scalars with a Weyl symmetry are not useful
to understand the solutions and their extension of the Einstein
domain. Actually, the path of discovery was to first obtain this
action from 2T-physics \cite{2Tgravity}\cite{2TgravityGeometry},
which then enabled us to find the solutions \cite{inflationBC}.}.

The usual Einstein gravity in Eq.(\ref{action}) can be obtained by
choosing a gauge for the local scale symmetry. In particular, in the
Einstein gauge we denote the fields by $\phi_{E}\left(  x\right)
,s_{E}\left(  x\right) ,g_{E}^{\mu\nu}\left(  x\right)  $ with a
subscript $E,$ and we choose $\frac{1}{12}\left(  \phi_{E}^{2}\left(
x\right)  -s_{E}^{2}\left(  x\right) \right)
=\frac{1}{2\kappa^{2}}$ for all spacetime $x^{\mu},$ thus generating
the gravitational constant. Then writing
\begin{equation}
\phi_{E}\left(  x\right)  =\pm\frac{\sqrt{6}}{\kappa}\cosh\left(
\frac {\kappa\sigma\left(  x\right)  }{\sqrt{6}}\right)
,\;s_{E}\left(  x\right) =\pm\frac{\sqrt{6}}{\kappa}\sinh\left(
\frac{\kappa\sigma\left(  x\right)
}{\sqrt{6}}\right)  ,\label{E-gauge}%
\end{equation}
we find that the conformally gauge invariant action (\ref{C-action})
yields the familiar action in Eq.(\ref{action}) with an
Einstein-Hilbert term $\frac{1}{2\kappa^{2}}R\left(  g_{E}\right)  $
and a minimally coupled scalar field $\sigma$, and any potential
energy $V\left(  \sigma\right)  .$

Another convenient gauge choice is the $\gamma$-gauge in which we
denote the fields by $\phi_{\gamma}\left(  x\right)
,s_{\gamma}\left(  x\right) ,g_{\gamma}^{\mu\nu}\left(  x\right)  $
with a subscript $\gamma$. In this gauge we choose the conformal
factor of the metric $a_{\gamma}^{2}\left( x^{\mu}\right)  =1$ for
all spacetime $\left\{  x^{\mu}\right\}  .$ In
particular, in the FRW, Kasner or Bianchi IX spacetimes of Eqs.(\ref{metric}%
-\ref{bianchi9}) taken in the gamma gauge $a_{\gamma}^{2}\left(
\tau\right) =1 $, we obtain simple expressions for the curvature,
such as $R\left( g_{\gamma}^{FRW}\right)  =6K$, etc.. Inserting
these expressions in the full
action in Eq.(\ref{C-action}) we derive the gauge fixed action $S_{\text{eff}%
}^{\gamma}$ for cosmological applications, precisely as given in
Eq.(\ref{SeffG}).

It is useful to point our that the following quantity is gauge
invariant in all spacetime
\[
z\left(  x^{\mu}\right)  =\frac{\kappa^{2}}{6}\left(  -g\left(
x\right) \right)  ^{\frac{1}{4}}\left(  \phi^{2}\left(  x\right)
-s^{2}\left( x\right)  \right) .
\]
In the Einstein gauge in the gravity sector given in
Eq.(\ref{E-gauge}) it is equal to $z\left(  x^{\mu}\right)  =\left(
-g\left(  x\right)  \right) ^{\frac{1}{4}}=a_{E}^{2}\left(  x\right)
$, while in the Einstein gauge in
the antigravity sector (interchanging $\cosh$ and $\sinh$ in Eq.(\ref{E-gauge}%
)) it is equal to $z\left(  x^{\mu}\right)  =-\left(  -g\left(
x\right) \right)  ^{\frac{1}{4}}=-a_{E}^{2}\left(  x\right)  .$
Furthermore, in the
$\gamma$-gauge it is equal to $z\left(  x^{\mu}\right)  =\frac{\kappa^{2}}%
{6}\left(  \phi_{\gamma}^{2}-s_{\gamma}^{2}\right)  $. Equating
these quantities to each other by gauge invariance immediately gives
part of the BCST transformation in Eq.(\ref{bcst1}). This shows that
the BCST transformation in Eqs.(\ref{bcst1},\ref{bcst2}) is actually
the gauge transformation from the Einstein gauge (both gravity and
antigravity regions) to the $\gamma$-gauge and vice-versa.

Now it is clear that the $\gamma$-gauge of the action
(\ref{C-action}) is valid in the entire $\left(  \phi_{\gamma}\left(
x\right)  ,s_{\gamma}\left( x\right)  \right)  $ plane in all
spacetime, for all values of the gauge invariant $z\left(  x\right)
$. But the usual Einstein theory as given in Eq.(\ref{action}) is
possible only in the left and right quadrants of Fig.1 where $\left(
\phi_{\gamma}^{2}-s_{\gamma}^{2}\right)  >0,$ or equivalently when
the gauge invariant $z\left(  x\right)  $ is positive. To give an
Einstein frame description of the top and bottom quadrants of Fig.1,
where $\left(  \phi^{2}-s^{2}\right)  <0$ and the gauge invariant
$z\left(
x\right)  $ is negative, we need an Einstein gauge just like Eq.(\ref{E-gauge}%
) by interchanging $\cosh$ with $\sinh.$ The result of gauge fixing
in the region $\left(  \phi^{2}-s^{2}\right)  <0$ is an Einstein
action just like (\ref{action}), but with the gravitational constant
$1/2\kappa^{2}$ replaced by $-1/2\kappa^{2},$ and the wrong sign
kinetic term for the sigma field. Evidently, in this system there is
antigravity rather than gravity since the Newton constant has the
opposite sign.

Comparing the gravity/antigravity regions further in their
respective Einstein frames, note that the potential energy $V\left(
\sigma\right)  $ does not change sign, but is replaced by a new one
$\bar{V}\left(  \sigma\right)  $ which is related to the old one in
Eq.(\ref{V}) by interchanging $\cosh$ with $\sinh.$ Any additional
terms in a complete action do not change sign either, but any terms
that initially contain $\phi,s$ would look different in the
gravity/antigravity sectors by the interchange of $\cosh$ with
$\sinh$. The $\gamma$-gauge interpolates smoothly between the
gravity and antigravity regions and this is why, among other things,
it is helpful with the geodesic completeness of the geometry.

After discussing the classical solutions of the cosmological
equations, we will return to the antigravity region to discuss
possible instabilities due to the sign changes in the full action
(\ref{C-action}), and what they may mean.

\section{Properties of the solutions}

First, recall the geodesic incompleteness of the geometry in the
usual Einstein frame, which is an old problem that has been set
aside. This is typical for any FRW cosmological solution with a big
bang singularity, because the geometry is incomplete at the
singularity as $a_{E}^{2}\left( \tau\right)  \rightarrow0$. A
massive particle in the flat FRW universe
satisfies the geodesic equation $d\vec{x}/d\tau=\vec{p}/\sqrt{\vec{p}%
^{2}+m^{2}a_{E}^{2}\left(  \tau\right)  }$ where $\vec{p}$ is the
conserved momentum of the particle. From this it is straightforward
to compute that it takes a finite amount of conformal time, cosmic
time, or proper time, to reach the big bang singularity from
anywhere in the universe. Therefore the particle trajectory is
artificially stopped at the singularity in a finite amount of proper
time. This is geodesic incompleteness. It begs the question: what
was there just before the big bang? What would we find if we could
allow our watch to run beyond the singularity where it was
artificially stopped? A geodesically complete geometry would provide
at least a partial answer to this question.

Of course, modifications of the geometry or even a completely
different description may be expected from quantum gravity or string
theory. Unfortunately, the question remains even more obscure in
these formalisms because of the lack of proper understanding of how
they should be applied to the cosmological setting. It is therefore
not unreasonable to at least, at first, try to answer such a
question in the setting of classical gravity. Having the complete
set of solutions analytically allowed us to answer such questions,
and understand how to complete the geometry in the classical physics
setting, thus leading to some surprising behavior, as follows.

The complete set of solutions of the cosmological equations have
been analyzed in detail in
\cite{inflationBC}\cite{BCT1}\cite{BCST2}. These solutions are
obtained with \textit{no restrictions on either the parameters of
the model or the initial conditions on the fields}. Before the
important role of anisotropy is taken into account near the
singularity at $a_{E}^{2}=0 $, there are 6 unrestricted parameters
in our solutions as indicated following Eq.(\ref{SeffG}). All 6
parameters are available to try to fit cosmological data far away
from the singularity.

In the absence of anisotropy, the \textit{generic solution} for
$\left( \phi_{\gamma}\left(  \tau\right)  ,s_{\gamma}\left(
\tau\right)  \right)  $ behaves in various detailed ways in various
regions of the 6 parameter space. This is given in the Appendix of
\cite{BCST2} where we identified 25 different regions of the 6
parameter space in which the analytic expression is different for
each case separately. The trajectory of the generic solution in the
absence of anisotropy can be plotted parameterically in the $\left(
\phi_{\gamma},s_{\gamma}\right)  $ plane of Fig.1. This shows that
the generic trajectory crosses the lightcone in Fig.1 at any point
(as determined by the 6
parameters) so that $\phi_{\gamma}^{2}\left(  \tau\right)  -s_{\gamma}%
^{2}\left(  \tau\right)  $ keeps changing sign as the universe moves
back and forth from the gravity patch to the antigravity patch. Of
course, to do so, the scale factor $a_{E}^{2}\left(  \tau\right)  $
vanishes at each crossing, so that the gravity/antigravity patches
are connected to each other only through the cosmological
singularity.

The behavior of the generic solution near the singularity changes
dramatically in the presence of anisotropy. As will be outlined
below, an important effect of anisotropy is to focus the trajectory
of the generic solution to pass through the origin of the $\left(
\phi_{\gamma},s_{\gamma}\right)  $ plane, such that the generic
trajectory can cross the lightcone in Fig.1 only at the origin where
$\left(  \phi_{\gamma},s_{\gamma}\right)  $ vanish simultaneously.

\subsection{Zero-size and Finite-size Bounces in the Gravity Patch}

By restricting the parameter space or initial conditions, we find
that there is a subset of special solutions that are geodesically
complete in the Einstein frame purely in the gravity patch. These
are the zero-size and finite-size bounces shown in Figs.(2,3,4,5).
As seen in Figs.(4,5), their trajectories never reach into the
antigravity sector and they never cross the lightcone in Fig.1
except at the origin.

\begin{center}
$%
\begin{array}
[c]{cc}%
{\parbox[b]{2.9006in}{\begin{center}
\includegraphics[
height=1.804in, width=2.9006in
]%
{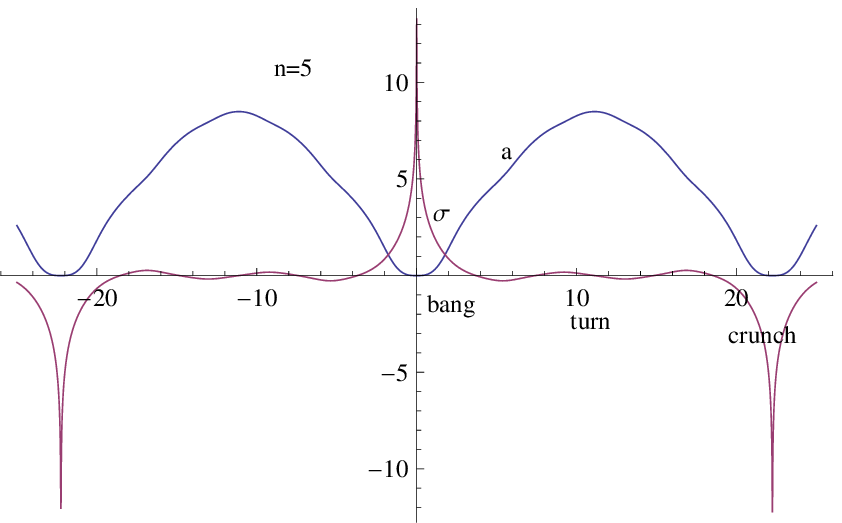}%
\\
Fig.2 - Zero-size bounces in a cyclic universe. $a_E\left(
\tau\right)  $ and $\sigma\left(  \tau\right)  $ for $b<0$ is
depicted. For $b>0$ turnaround is at $a_E\rightarrow\infty.$
\end{center}}}%
&
{\parbox[b]{2.8928in}{\begin{center}
\includegraphics[
height=1.7997in, width=2.8928in
]%
{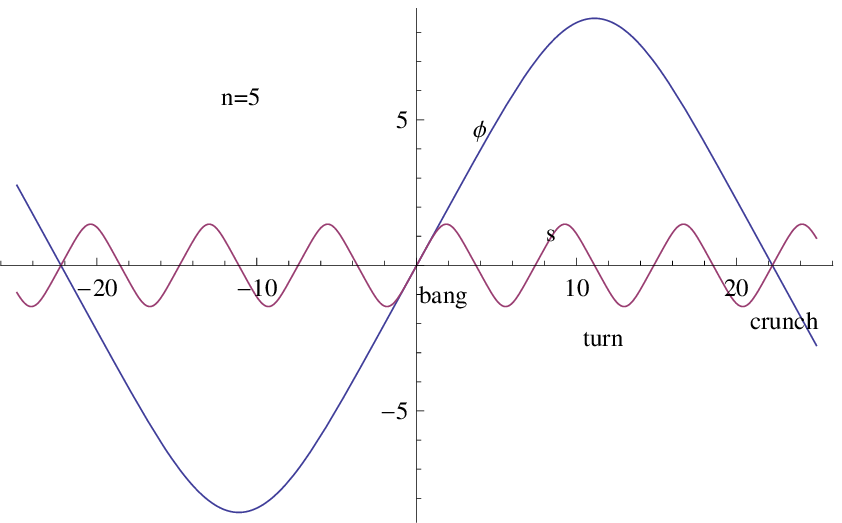}%
\\
Fig.3 - $\phi\left(  \tau\right)  ,s\left(  \tau\right)  $ for
$b<0,$ with synchronized initial values and relatively quantized
periods. For $b>0$ turnaround is at $\phi\rightarrow\infty.$
\end{center}}}%
\end{array}
$
\end{center}

The zero-size-bounce (see Figs.(2,3)) describes a cyclic universe
that \textit{contracts to zero size and then bounces back smoothly
from zero size}. It expands up to either a finite size (when $b<0$)
or infinite size (when $b>0$) and then turns around to repeat the
cycle. In these solutions, as
described in detail in \cite{BCT1}\cite{BCST2}, the behavior of $a_{E}%
^{2}\left(  \tau\right)  $ near the singularity at $\tau\sim0$ is
smooth (if $\rho_{0}$ or $K$ take generic values, then
$a_{E}^{2}\left(  \tau\right) \sim\tau^{2}\rightarrow0$; if both
$\rho_{0},K$ vanish or take some special values, then
$a_{E}^{2}\left(  \tau\right)  \sim\tau^{6}\rightarrow0).$ Also, the
behavior of the potential and kinetic energy terms for the scalar
field $\sigma\left(  \tau\right)  $ near the singularity are
surprising. Namely, contrary to the generic solution, the potential
energy dominates over the kinetic energy so that the equation of
state $w\left(  \tau\right)  $ is negative near the singularity.
According to common lore, it was thought that such zero-size-bounce
solutions would not exist because they would violate the null energy
condition (NEC). However, this is not the case. The NEC is satisfied
because there is a singularity at zero size, and this is the
exception allowed according to the NEC theorems. We found all such
solutions and classified them in \cite{BCT1}\cite{BCST2}, thus
providing a rich class of examples of zero-size-bounce cyclic
universes, characterized by arbitrary values of the parameters
$\rho_{0},K,b,c$ plus one additional quantized parameter.

To make the zero-size-bounce happen, a synchronization of initial
conditions and a quantization condition among the 6 available
parameters must be imposed. These properties are illustrated in
Fig.3, where it is seen $s_{\gamma}\left( 0\right)
=\phi_{\gamma}\left(  0\right)  =0$ is imposed as an initial
condition, and the periods of oscillations of $\phi\left(
\tau\right)  $ and $s\left(  \tau\right)  $ are quantized relative
to each other. Hence such solutions are characterized by 4
continuous and 1 quantized parameter, rather than the 6 continuous
parameters of the generic solutions. In that sense the
zero-size-bounce solutions are a set of measure zero. So,
statistically, it does not seem likely that the universe would
choose such a solution over the generic solution.

\begin{center}
$%
\begin{array}
[c]{cc}%
{\parbox[b]{2.9144in}{\begin{center}
\includegraphics[
height=0.5076in, width=2.9144in
]%
{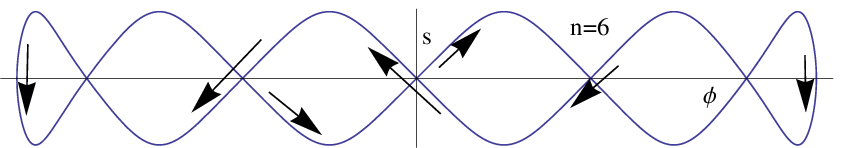}%
\\
Fig.4 - Geodesically complete trajectory only in the gravity sector,
for $b<0.$
\end{center}}}%
&
{\parbox[b]{3.2059in}{\begin{center}
\includegraphics[
height=0.2741in, width=3.2059in
]%
{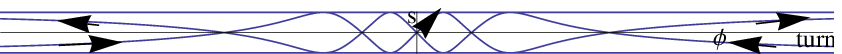}%
\\
Fig.5 - Geodesically complete trajectory only in the gravity sector,
for $b>0.$
\end{center}}}%
\end{array}
$
\end{center}

However, anisotropy plays a very important role in providing an
attractor mechanism such that, for typical initial conditions away
from the singularity, all trajectories are attracted to the origin
of the $\left(  \phi,s\right)  $ plane and can cross the lightcone
in Fig.1 only at the origin. In that sense the type of solution
depicted in Figs.(2-5) is not too far from being generic once
anisotropy is taken into account. In particular the behavior away
from the origin is a good approximation. Nevertheless, what goes on
in the singularity region is quite different as seen in Figs.(8 ,9)
and discussed below.

\begin{center}
$%
\begin{array}
[c]{cc}%
{\parbox[b]{2.8651in}{\begin{center}
\includegraphics[
height=1.7824in, width=2.8651in
]%
{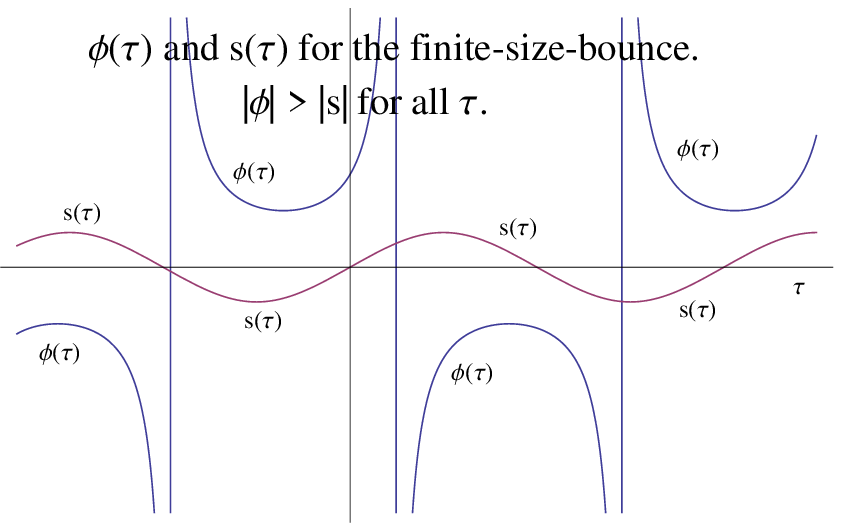}%
\\
Fig.6 - Finite-size-bounce, $\left\vert \phi_\text{min}\left(
\tau\right) \right\vert >K/4b>\left\vert s_\text{max}\left(
\tau\right)  \right\vert .$
\end{center}}}%
&
{\parbox[b]{2.9006in}{\begin{center}
\includegraphics[
height=1.804in, width=2.9006in
]%
{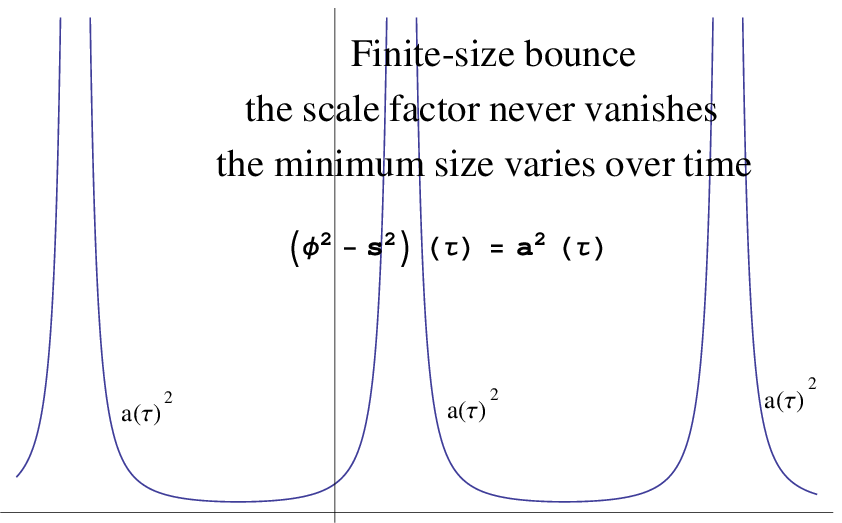}%
\\
Fig.7 - Finite-size-bounce, \\[0pt]$a_E^2\left(  \tau\right)  $ never
vanishes.
\end{center}}}%
\end{array}
$
\end{center}

The finite-size-bounce (see Figs.(6,7)) describes a universe that
contracts up to a minimum non-zero size and then bounces back into
an expansion phase up to infinite size. As the universe turns around
to repeat such cycles the minimum size is not necessarily the same,
as this depends on the parameters. Such solutions occur when the
parameters satisfy, $\rho_{0}<K^{2}/16b$ and
$\phi_{\text{min}}^{2}\left(  \tau_{0}\right)
>K/4b>s_{\text{max}}^{2}\left( \tau_{0}\right)  $. Note that there
are still 6 parameters, so this is not a set of measure zero, but it
is a restricted region of parameter space or
initial values. The analytic solution is given explicitly in \cite{BCT1}%
\cite{BCST2}.

\section{Anisotropy and the Attractor}

It is useful to write the equations of motion for the fields $\sigma
,\alpha_{1},\alpha_{2}$ in terms of the canonical momenta instead of
their
velocities. These are given by $z\dot{\sigma}=p_{\sigma}$ , $z\dot{\alpha}%
_{1}=p_{1}$ and $z\dot{\alpha}_{2}=p_{2}.$ Then the energy
constraint that follows from the action $S_{\text{eff}}^{\gamma}$ or
$S_{\text{eff}}^{E}$ takes the following form when written in terms
of $z\left(  \tau\right)
=\phi_{\gamma}^{2}\left(  \tau\right)  -s_{\gamma}^{2}\left(  \tau\right)  $%
\begin{equation}
\frac{\dot{z}^{2}}{4z^{3}}=\frac{\kappa^{2}}{3}\left[  \frac{p_{\sigma}%
^{2}+p_{\alpha}^{2}}{2z^{3}}+V\left(  \sigma\right)  +\frac{\rho_{0}}{z^{2}%
}\right]  -\frac{6Kv\left(  \alpha_{1},\alpha_{2}\right)  }{\kappa^{2}%
z},\label{adot}%
\end{equation}
where $p_{\alpha}^{2}\equiv p_{1}^{2}+p_{2}^{2}.$ For the generic
solution (i.e. unlike the zero-size-bounce solution, assuming
typically $(p_{\sigma },\ p_{1},\ p_{1}) $ are non-vanishing at the
singularity), the kinetic term
$\frac{p_{\sigma}^{2}+p_{\alpha}^{2}}{2z^{3}}$ is the dominant term
in the energy equation as we approach the singularity
$z\rightarrow0.$ The next to the leading term is
$\frac{\rho_{0}}{z^{2}}$ while the potential terms $V\left(
\sigma\right)  ,\frac{Kv\left(  \alpha_{1},\alpha_{2}\right)  }{z}$
are subdominant. Keeping the leading and next to the leading terms,
it is possible to integrate all the Friedmann equations to obtain
the behavior of the general solution near the singularity. In the
absence of the potentials, the momenta $\left(
p_{\sigma},p_{1},p_{2}\right)  $ are all conserved; then the
solution is uniquely given as follows (where we use units with
$\kappa=\sqrt{6})$
\begin{equation}
z\left(  \tau\right)  =2\tau\left(  \sqrt{p_{\sigma}^{2}+p_{\alpha}^{2}}%
+\rho_{0}\tau\right)  ,\;\;a_{E}^{2}\left(  \tau\right)  =\left\vert
z\left(
\tau\right)  \right\vert , \qquad\sigma\left(  \tau\right)  =\frac{p_{\sigma}%
}{4\sqrt{p_{\sigma}^{2}+p_{\alpha}^{2}}}\ln\left(  \frac{\left(
\tau/T\right)  ^{2}}{\left(  \sqrt{p_{\sigma}^{2}+p_{\alpha}^{2}}+\rho_{0}%
\tau\right)  ^{2}}\right)  ,\label{z}%
\end{equation}
as well as
\begin{equation}
\phi_{\gamma}\left(  \tau\right)  +s_{\gamma}\left(  \tau\right)
=\sqrt {T}\left(
\sqrt{p_{\sigma}^{2}+p_{\alpha}^{2}}+\rho_{0}\tau\right)  \left(
\frac{\left(  \tau/T\right)  ^{2}}{\left(  \sqrt{p_{\sigma}^{2}+p_{\alpha}%
^{2}}+\rho_{0}\tau\right)  ^{2}}\right)  ^{\frac{1}{4}\left(
1+\frac
{p_{\sigma}}{\sqrt{p_{\sigma}^{2}+p_{\alpha}^{2}}}\right)  },\label{f+s}%
\end{equation}
\begin{equation}
\phi_{\gamma}\left(  \tau\right)  -s_{\gamma}\left(  \tau\right)
=\frac {2\tau}{\sqrt{T}}\left(  \frac{\left(  \tau/T\right)
^{2}}{\left(
\sqrt{p_{\sigma}^{2}+p_{\alpha}^{2}}+\rho_{0}\tau\right)
^{2}}\right)
^{-\frac{1}{4}\left(  1+\frac{p_{\sigma}}{\sqrt{p_{\sigma}^{2}+p_{\alpha}^{2}%
}}\right)  }.\label{f-s}%
\end{equation}
where $T$ is an integration constant that amounts to an initial
value for $\sigma$. Similarly, the solutions for $\alpha_{1}\left(
\tau\right) ,\alpha _{2}\left( \tau\right)  $ near the singularity
are
\begin{equation}
\alpha_{i}\left(  \tau\right)
=\frac{p_{i}}{4\sqrt{p_{\sigma}^{2}+p_{\alpha }^{2}}}\ln\left(
\frac{\left(  \tau/T_i\right)  ^{2}}{\left(  \sqrt{p_{\sigma
}^{2}+p_{\alpha}^{2}}+\rho_{0}\tau\right)  ^{2}}\right)  ,\;i=1,2.
\end{equation}
where $T_i$ are integration constants that amount to initial values
for the $\alpha_i$.

The solutions for $\left(  z\left(  \tau\right)  ,\sigma\left(
\tau\right) \right)  $ are plotted in Fig.8, where we see that
$z\left(  \tau\right) =\phi_{\gamma}^{2}\left(  \tau\right)
-s_{\gamma}^{2}\left(  \tau\right)  $ changes sign. This shows that
the trajectory of the universe inevitably passes through
antigravity. Furthermore, the parametric plot in Fig.9 shows that
there is an attractor mechanism that forces the trajectory to pass
through the origin of the $\left(  \phi_{\gamma},s_{\gamma}\right)
$ plane.

\begin{center}
$%
\begin{array}
[c]{cc}%
{\parbox[b]{3.0251in}{\begin{center}
\includegraphics[
height=1.881in, width=3.0251in
]%
{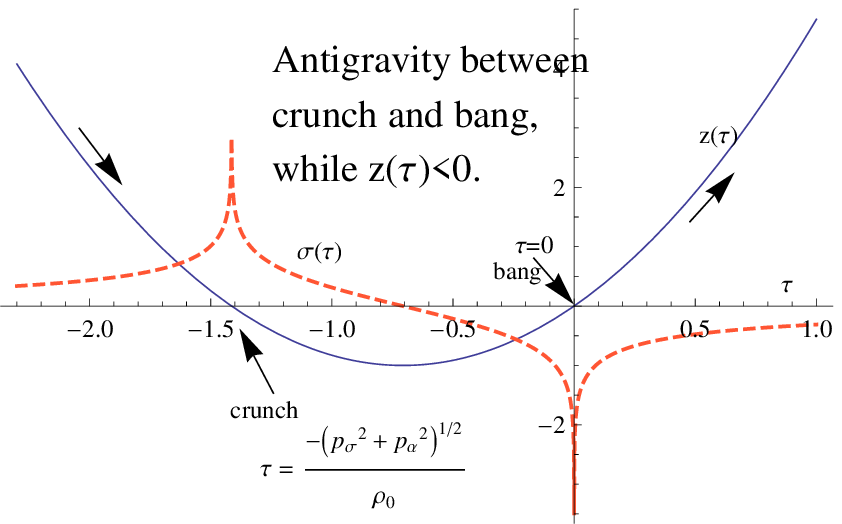}%
\\
Fig.8 - Plots of $z\left(  \tau\right)  $ and $\sigma\left(
\tau\right)  ,$ with $all$ parameters $\rho_0,p_\sigma,p_\alpha,T,$
set to 1 numerically.
\end{center}}}%
&
\raisebox{-0pt}{\parbox[b]{3.2128in}{\begin{center}
\includegraphics[
height=1.8922in, width=3.2128in
]%
{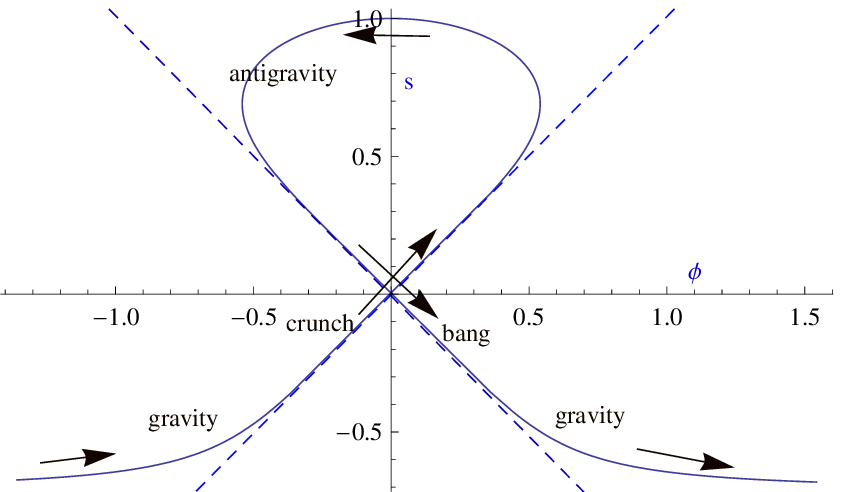}%
\\
Fig.9 - Antigravity loop in the parametric plot of $\phi\left(
\tau\right) ,s\left(  \tau\right)  .$
\end{center}}}%
\end{array}
$
\end{center}

From these plots it is seen that the universe contracts to zero size
with a big crunch, passing through the singularity at $z\left(
-\tau_{c}\right) =0$, it then expands in a region of antigravity
while $z\left(  \tau\right) <0,$ then contracts again to zero size
at $z\left(  0\right)  =0$, and then emerges with a big bang in a
region of gravity where $z\left(  \tau\right)
>0.$ The duration of the antigravity phase is
\begin{equation}
\tau_{c}=\frac{\sqrt{p_{\sigma}^{2}+p_{\alpha}^{2}}}{\rho_{0}}%
.\label{duration}%
\end{equation}
The essential parameters that control the properties of this unique
solution
in the vicinity of the singularity are $p_{\sigma},\sqrt{p_{1}^{2}+p_{2}^{2}%
},\rho_{0}.$ They are related to the kinetic energy of the sigma
field, anisotropy and \textquotedblleft radiation\textquotedblright\
(i.e. all relativistic particles in the early universe)
respectively. They are all typically non-zero and not negligible in
the vicinity of the singularity. Note that \textit{even if
anisotropy }$\sqrt{p_{1}^{2}+p_{2}^{2}}$ \textit{is infinitesimal,
the attraction to the origin } $\phi=s=0$\textit{\ is inevitable}.

The parameter space for which the potentials can be neglected is
estimated by inserting the solution back into the the equations and
comparing the magnitude of the potential energy terms to the kinetic
energy and radiation terms. From this we determined that the
radiation term $\rho_{0}/z^{2}\left(  \tau\right) $ in
Eq.(\ref{adot}) beats the potential $V\left(  \sigma\right)  $ by
more than two powers of $\tau$ as $\tau\rightarrow0,$ and similarly
for $\tau\rightarrow-\tau_{c},$ so $V\left(  \sigma\left(
\tau\right)  \right)  $ is negligible near the singularity.
Similarly $Kv\left(  \alpha_{1},\alpha _{2}\right)  /z$ is
negligible compared to radiation as a function of time
\textit{near} the singularity (even when $K\neq$0) provided $p_{\sigma}%
^{2}>15\left(  p_{1}^{2}+p_{2}^{2}\right)  $. This condition is also
compatible with a weaker condition ($p_{\sigma}^{2}>3\left(  p_{1}^{2}%
+p_{2}^{2}\right)  $, see \cite{BCST2}) that avoids the mixmaster
behavior discussed in \cite{Misner} and is consistent with the
discussions in \cite{BKL}\cite{Damour}.

The attractor mechanism is an almost model independent universal
behavior since the potential energy $V\left(  \sigma\right)  $ is
negligible close to the singularity. Anisotropy in even tiny amounts
forces the universe to behave as described above. The issue of
\textit{initial conditions for all degrees of freedom in our model
becomes partially resolved} because at the singularity $z=0$ we must
have the following automatically enforced initial values,
independent of any other arbitrarily chosen initial conditions at
some other time,
\begin{equation}
\phi_{\gamma}=s_{\gamma}=0,\label{initial}%
\end{equation}
rather than only $\phi_{\gamma}^{2}-s_{\gamma}^{2}=z=0$. Hence, we
must take $\phi_{\gamma}\left(  0\right)  =s_{\gamma}\left(
0\right)  =0$ as the dynamically enforced boundary values at all
cosmological singularities. This reduces the number of arbitrary
initial values.

Now we see that the zero-size-bounce solutions described in the
previous section (Figs.(2-5)) are not really of measure zero. They
do take into account effectively the dynamically enforced initial
values of Eq.(\ref{initial}). Although these solutions are not
capable of describing the detailed (antigravity loop) behavior near
the singularity, they do describe the correct analytic behavior away
from the singularity where anisotropy becomes negligibly small. A
more accurate geodesically complete solution is the combination of
the near-singularity and far-from-singularity solutions which can be
constructed by matching them at some time sufficiently away from the
singularity.

\section{Antigravity Regime}

\label{antigrav}

The physics during the antigravity period remains cloudy in our
investigation so far. We discussed the trajectory of the universe on
the average, including the antigravity period. However, once in this
regime, we may expect some violent behavior and backreaction because
of antigravity, and these may affect the trajectory during the
antigravity period. First, because of the opposite sign of the
gravitational constant we expect that chunks of similar matter will
repel each other rather than attract. This is an invitation to
investigate the nature of interactions, including fluctuations on
top of our background solution, which so far we have not included in
the analysis. Furthermore, there are additional phenomena we should
expect due to interactions as seen by investigating the form of the
action in the antigravity regime, as follows.

Because of the sign change of the effective Newton constant one may
argue that fluctuations of the metric, namely the spin-2 gravitons,
will now come with the wrong sign kinetic terms. There are no gauge
symmetries to remove them, hence, they will behave like negative
energy fluctuations, which could destabilize the vacuum by the
emission of negative energy gravitons together with lots of positive
energy matter particles. There is however a counter effect. Once
particle production begins, it automatically implies a back reaction
that changes the energy density $\rho$ for all relativistic matter
that enters in the cosmological equations. This effectively amounts
to a huge increase of the parameter $\rho_{0}$ once we enter the
antigravity regime. So, rather than solving the cosmological
equations with a constant $\rho_{0}$ in the antigravity region, we
should allow $\rho_{0}$ to increase dramatically by large amounts.
Then according to our rough formula in Eq.(\ref{duration}) for the
duration of the antigravity regime, the duration gets shortened as
$\rho_{0}$ increases. In other words, particle production by
antigravity shortens its duration. The particle production and the
corresponding jump of $\rho_{0}$ need to be better understood. These
are effects that oppose each other. It is difficult to estimate the
balance of these effects with our current understanding of the
corresponding physics.

Of course, quantum gravity effects need to be also included.
Therefore, it would be very interesting to study similar
circumstances in the framework of string theory, although the
requisite technical tools may not be available at this time. To
formulate the antigravity aspects in string theory we could use the
BCST field transformations given in Eqs.(\ref{bcst1},\ref{bcst2}),
but even better would be the inclusion of the analog of the Weyl
symmetry in the framework of string theory.

We should also mention that the phenomenon of the sign change of the
Newton constant transcends the specific simple model in
Eq.(\ref{C-action}). The phenomena we have found should also be
expected generically in supergravity theories coupled to matter
whose formulation include a similar factor that multiplies $R\left(
g\right)  .$ In supergravity, that factor is related to what is
called the Kahler potential, and this factor, combined with the
usual Einstein-Hilbert term, is not generally positive definite
\cite{weinberg}. In the past, typically it was assumed that the
overall factor is positive and investigations of supergravity
proceeded only in the positive regime. A
discussion of the field space in the positive sector for general $\mathcal{N}%
$=2 supergravity can be found in \cite{deWit}. However, our results
suggest that generically the overall factor can and will change sign
dynamically, and therefore antigravity sectors similar to our
discussion in this paper should be expected in typical supergravity
theories. Furthermore, we can also draw an analogy to similar
factors that occur in front of the kinetic terms for gauge bosons
and scalar fields \cite{weinberg}. These too can potentially change
sign. In the case of gauge bosons, since the kinetic term takes the
form $\left(  E^{2}-B^{2}\right)  $ where $E,B$ represent electric
and magnetic fields, a sign change could be reinterpreted by an
electric-magnetic duality transformation. Indeed this type of
duality was explored in depth at the quantum level for
$\mathcal{N}$=2 super Yang-Mills theory in \cite{SW}. It is
therefore interesting to point out that something similar may be
part of the phenomena in the case of the gravity/antigravity
transitions we have uncovered.

Having mentioned a few of the possible physical effects in the
antigravity regime, we will proceed by assuming that the
destabilizing effects of antigravity are not a disaster, but rather
they are the signals of interesting physics that remains to be
understood better. We expect that the general nature of our solution
survives and still describes the interesting effects we discovered,
namely the attractor mechanism and the presence of antigravity for
some period of time between a crunch and the following bang.

\section{Summary}

\label{discussion}

In this talk I emphasized the role of analytic solutions of
cosmological equations to discuss geodesic completeness through the
big bang singularity and the discovery of some new phenomena by
focussing on questions that could not be answered with only
approximate solutions. Until better understood in the context of
quantum gravity, or string theory, our results should be considered
to be a first pass for the types of new questions they raise and the
answers they provide. Here is a brief summary of the main points we
have learned.

\begin{itemize}
\item We found new techniques to solve cosmological equations analytically.
This led to all the solutions for several special potentials
$V(\sigma)$ (only one of those was discussed in this talk). Several
model independent general results that followed from this include
geodesic completeness, and an attractor mechanism to the origin of
the $\left(  \phi_{\gamma},s_{\gamma }\right)  $ plane, which
resolves partially questions on initial values.

\item Antigravity is very hard to avoid. Anisotropy + radiation + kinetic
energy of $\sigma,$ require the antigravity epoch. For close to a
year we tried to find models and mechanisms to avoid antigravity
(i.e. when \textit{all} solutions of a geodesically complete model
are included). The failure to find such mechanisms finally led us to
take antigravity seriously.

\item We studied the Wheeler-deWitt equation to take into account some quantum
effects for the same system that we analyzed classically. We could
solve some cases exactly, others semi-classically. This will be
published at a later time. But let me mention that we arrived
substantially to the same conclusions that we learned by studying
the purely classical system.

\item Will the new insights we obtained survive the effects of a full quantum
theory? My expectation is affirmative. Perhaps the most satisfactory
approach is to try to study the same issues in the context of string
theory. However, this is actually quite challenging since string
theory has not been developed sufficiently in the cosmological
context.

\item I emphasize that the structure of the theory in Eq.(\ref{C-action}), the
methods of solutions, and the phenomena that ensued, are
historically direct
predictions of 2T-physics in 4+2 dimensions \cite{2Tgravity}%
\cite{2TgravityGeometry}. The model corresponds to the conformal
shadow of 2T-physics which takes the form of a familiar 1T-field
theory setting. Other shadows \cite{2TPhaseSpace} of the same 2T
theory which would be in the form of dual 1T-field theories, could
also play an important role to understand the system and its
physical interpretation.

\item Open questions include: are there observational effects today of a past
antigravity period? This is an important ongoing project for our
group. This involves the study of small fluctuations and the fitting
to current observations of the CMB.
\end{itemize}

Much remains to be understood, including quantum gravity and string
theory effects, but it is clear that previously unsuspected
phenomena, including an attractor mechanism and antigravity, are at
work close to the cosmological singularity. The technical tools to
study such issues in the context of a full quantum theory of gravity
are yet to be developed. So this is an important challenge to the
theory community, including string theorists, since here are some
deep physics issues where string theory, if it is the correct
approach at those scales, should play its most important role as a
quantum theory of gravity. Perhaps duality transformations of the
type that occur in electric-magnetic duality \cite{SW} could play a
role as mentioned earlier. Our understanding of cosmology and of the
beginning of the universe could not be clarified until an
understanding of the antigravity effects are obtained in a full
quantum theory of gravity, and the associated phenomena are taken
into account in comparing to cosmological observations.



\begin{acknowledgments}
This research was partially supported by the U.S. Department of
Energy under grant number DE-FG03-84ER40168. It was also partially
supported by Princeton University and the Perimeter Institute during
my sabbatical at those institutions. I would like to thank my
collaborators C.H. Chen, Paul Steinhardt and Neil Turok with whom
this research was conducted. I also thank Niayesh Afshordi, Tom
Banks, Latham Boyle, Bernard deWit, and Martin Rocek for helpful
conversations.
\end{acknowledgments}

\bigskip

\end{document}